# An Overview of Computer Security

By
**Shireesh Reddy Annam**

Undergraduate, Department of Electrical Engineering

Indian Institute of Technology Kanpur, Kanpur, Uttar Pradesh, India 208016

## ABSTRACT

As more business activities are being automated and an increasing number of computers are being used to store vital and sensitive information the need for secure computer systems becomes more apparent. These systems can be achieved only through systematic design; they cannot be achieved through haphazard seat-of-the-pants methods.

This paper introduces some known threats to the computer security, categorizes the threats, and analyses protection mechanisms and techniques for countering the threats. The threats have been classified more so as definitions and then followed by the classifications of these threats. Also mentioned are the protection mechanisms.

1. **INTRODUCTION**

Computer security consists largely of defensive methods used to detect and thwart would-be intruders. The principles of computer security thus arise from the kinds of threats intruders can impose. This paper begins by giving examples of known security threats in existing systems. The second section presents a classification of security threats, and the last section presents some protection mechanisms and techniques for ensuring security of a computer system. This paper doesn't address the topics of physical security, communication security, and breaches of trust by personnel with the access to sensitive information.

## 2. THREATS

Probably the most publicized threat is the result of an intruder guessing a user's password. With the advent of personal computers, dial-up modems and proxy servers this has become much more of a problem. Penetrators have a list of the commonly used passwords and they can then try them all with the aid of their personal computer. In addition, if passwords are short they are easily found by an exhaustive search [**1**]. There are also standard accounts with default passwords that are distributed with systems, and may not have been changed.

Another common threat is called *spoofing*. This is accomplished by fooling a user into believing that he/she is talking to the system, resulting in information being revealed. For instance, a spoofer can make an unsuspecting user accesses the *web* funneled through the spoofer's machine, allowing the spoofer to monitor all of the victim's activities including any passwords or account numbers the victim enters. The spoofer can also cause false or misleading data to be sent to web servers in the victim's name, or to the victim in the name of any web server. In short, the attacker observes and controls everything the victim does on the web. Another example that can be cited for instance, the spoofer may display what looks like the system login prompt on a terminal to make the terminal appear to be idle. Then when an unsuspecting user begins to communicate with the terminal, the spoofer retrieves the login name and asks for the user's password. After obtaining the information, the spoofer displays a try again message or something else and returns ownership that was previously obtained by him.

Another threat is user browsing for sensitive information. This occurs when a legitimate user peruses any files that are available and gleans useful information. For instance, a browser may locate a password inadvertently left in a publicly readable file.

A more sophisticated threat, commonly known as the *Trojan horse*, is the result of a program doing more that it is supposed to or it's a program that appears to do something good, while it's actually doing something nasty in the background. For instance, a backgammon program or some software may be made public. However, when the unsuspecting plays against the program, the Trojan horse, executing with the user's own access rights to his files surreptitiously reads the user's files and might even also mail them to the creator of the game, if the user has himself logged onto the net. Recently, we saw reputed internet sites susceptible to what are called distributed denial of service attacks. Hackers using Trojans mastermind these attacks.

Another threat is the result of a devious user exhausting a shared resource so that legitimate users cannot complete the work. For instance, the devious user of a network front-end might use all of the available message buffers making it impossible for the legitimate users to accomplish any useful work. The intentional crashing of the system causing all work to a halt is a further example of this type of threat.

Another class of threats is the result of a user of a statistical database being able to infer sensitive data from non-sensitive information returned by the database. For instance, if Ram is the only psychology major in a particular class one can deduce Ram's grade from the average grade of the course and the average grade of all non-psychology majors in the class.

The reader may be asking , "If there is so much computer crime why haven't I heared about it?" Statistics show that approximately 1% of all computer crime is detected , 7% of the detected crrimes are reproted , 1 out of 33 criminals reported are convicted, and 1 out of 22,000 ends up in jail [**3**]. One reason crimes are not reported is that a successful attack often reveals vulnerabilities that can be exploited by other potential attackers. Furthermore, may of the crimes are viewed as pranks, and the people who detect them do not think they are serious enough to report to the police [**4**].

3. **THREAT CLASSIFICATION**

This section attempts to categorise the various threats. The classifications used were first used by Denning [**2**].

*Browsing* describes the method of searching through main and secondary memory for residue information. The browser is usually not looking for anything in particular, but is alert to possibly useful information. The breowser may find files containing sensitive information or containing information that helps to access other sensisitive information. The most useful deterrent to browsing is the use of controls that restrict users to only accessing information in their own data space.Enciphering data also deters browsing.

*Leakage* is the transmission of information to an unauthorised user from a process that is allowed to access the data. The public backgammon game is this type of threat.

An *inference* threat exists if a user can deduce sensitive information from non-sensitive data. This is usually the result of correlating information about groups of individuals to obtain information about an individual. The inference controls presented in the next section are used to counter this type of threat.

*Tampering* refers to the process of making unauthorised changes to the value of information stored in the computer. An example of tampering is a student changing his/her grade in the grade file. Tampering is avoided by allowing users to modify only their files. *Cryptography check summing* can be used for detecting tampering. This method uses cryptographic techniques, such as cipher block chaining, to generate a check sum for each file. The technique only detects changes; it doesn't prevent them.

*Accidental data destruction* although often innocent, can be costly. Accidental destruction may be caused by both hardware and software failures. For instance, faulty software could allow a program to write beyond its data space and overwrite another user's data. Access control techniques can be used to limit overwriting to the user's own data space. Cryptographic check summing can also be used for detecting accidental data destruction.

Browsing, leakage and inference are threats to the secrecy of data, and tampering and accidental destructions are threats to the integrity of data. Two threat classifications that fit into neither the secrecy or integrity category are masquerading and denial of service. *Masquerading* refers to the process where an intruder gains access to the stystem under another user's account. Spoofing and pasword guessing are masquerading threats. In the first the intruder is posing as the system, and in the second the intruder is posing as a legitimate user.

*Denial of service* threats prevent legitimate users from getting useful work done. The devious user exhausting all available resources is an example of this threat.

4. **PROTECTION MECHANISMS**

This section intoduces protection mechnasims used to enhance computer security. The mechnasims presented are grouped into authentication mechnasims, access control, and inference control. In additon, the methods of penetration analysis, formal verification techniques, and convert channel analysis are introduced.

*Authentication Mechanisms* – Authetication mechanisms primarily address the masquerading threat. The first mechanism discussed is the *secure attention key*. This key, when hit by a user at a terminal, kills any process running at the terminal except the true system listener and thus guarantees a trusted path to the system. This will foil attempts at spoofing the unsuspecting user. However, it is important that users make a habit of always hitting the secure attention key to begin a dialogue with the system. One way of ensuring this for the system to only display the login prompt after the key is depressed.

Simple guidelines can be used to deter password guessing. One should choose a long password (at least 8 characters) that is not obvious, and should not use easily guessable passwords like a spouse's name or a login name. In addition, a password should not be written in the obvious place. Furthermore, users should be trained to change their passwords at appropriate intervals. Most of the guidelines can be enforced by the system. For instance, password program can require long passwords and can check the password chosen against a dictionary of obvious passwords or something like reporting an error message if the password is the same as the login (a common practise by an average pc user). The login program can also inform the user that it is time to change passwords.

Password files stored in the system may be compromised like any other file. Therefore, it is not good practise to store passwords in the clear. Instead, a *one way function* (i.e., a function whose inverse is computationally infeasible to determine) is used to enchiper passwords and the result is stored in the password file. When a user's password is presented at the login time it is enchipered and compared to the stored value. By using one way functions to enchiper passwords the login file can be made public.

*Access Control* – Assuming that by using authentication mechanisms and good password practice the system can guarantee that users are who they claim to be, the next step is to provide a means of limiting a user's access to only those files that policy determines should be accessed. These controls are referred to as *access control*.

When describing access control policies and mechanisms it is necessary to consider the *subjects* and *objects* of the system. Subjects are the users of the system along with any active entities that act on behalf of the user or the system (eg. user processes). Objects are the resources or entities of the system (eg. files, programs, devices). The access control mechanism determines for each subject what *access modes* such as *read* (R), *write* (W), or *execute* (X), it has for each object.

A convenient way of describing a protection system is with an access matrix . In the access matrix rows correspond to subjects and columns correspond to objects. Each enrty in the matrix is a set of access rights that indicate the access that the subject associated with the row has for the object associated with the column. The following is an example access matrix. From the matrix one can determine that subject $S_3$ has read and write access to the object $O_2$ and execute access to the object $O_3$.

An example of access matrix

|  | OBJECTS | | | | |
| --- | --- | --- | --- | --- | --- |
| SUBJECTS | $O_1$ | $O_2$ | $O_3$ | $O_4$ | $O_5$ |
| $S_1$ | R |  | W | RW | W |
| $S_2$ |  | X |  | R |  |
| $S_3$ |  | RW | X |  |  |
| $S_4$ | RX |  | RW |  | RX |

There are two common ways of representing an access matrix in a computer system: access control lists (sometimes called authorization lists) and capability lists (often called c-lists). With the access list approach each object has an access list associated with it. This list contains the name of each subject that has access to the object along with the modes of access allowed. In contrast the capability list approach associates a list with each subject. The elements of the list are capabilities which can be thought of as tickets that contain an objects name and the modes of access allowed to the object. A subject's capability defines the environment or domain that the subject may directly access.

The reader should note that an access list corresponds to a column in the access and a capability corresponds to a row. An important aspect of either approach is that both the capabilities and the elements of access must be unforgeable or else the entire protection mechanism breaks down. One way of guaranteeing the unforgeability of these elements is by restricting access to them through an intermediatry trusted piece of code. The reference monitor introduced below is one such mechanism. Access control policies enforced by the access control mechanisms often incorporate *access hierarchies*. That is, subjects may have different ranks ranging from the most to the least privileged, where the more privileged user automatically gets the rights of the least privileged user. For instance, in a UNIX system a subject with the superuser privilege can access any object in the system.

As an example of an access control policy that incorporates access hierarchies, consider the *mandatory control policy*. In this model every subject and every object has an access class made up of a level (eg. unclassified, confidential, and secret) and a (possibly empty)set of categories (eg. crypto, nuclear, and intelligence). Levels are ordered and categories are not. When comparing access classes, the result can be say equal, less than, greater than , and not comparable. For instance, the access classes with level secret and category set containing only crypto is greater than the access class with the level unclassified and an empty category set. Furthermore, secret/{crypto} is less than secret/{crypto, nuclear}, and secret/{crypto} is not comparable to confidential/{nuclear}. The accesss rules for this policy are as follows. A subject may obtain read permission to an object if its access class is greater than or equal to the access class of the object. This is known as *simple security property*. In addition, a subject may write an object if the subject's access class is less than or equal to the access class of the object. This is known as *
property*.

To assure that all access control policies are enforced a means of mediating each access of an object by a subject is needed. The *reference monitor* provides this mediation. A reference monitor has three basic properties.
1. It must be tamperproof. That is, it should be isolated from modification by system entities.
2. It must always be invoked. That is, it must mediate every access.

3. It must be small enough to be subjected to analysis and tests, the completeness of which can be assured.

Reference monitors are often called *security kernels* in the literature.

*Inference Controls* – The last class of security mechanisms in this paper is the inference controls. These controls attempt to restrict access to sensitive information about individuals while providing access to statistics about groups of individuals. The ideal one is a statistical database that discloses no sensitive data.

Two approaches to sloving this problem are to restrict queries that reveal certain types of statistics and to add noise to the results returned. To foil small and large query set attacks, such as the Ram example in the first section, a technique called query-set-size control is introduced. This forbids the release of any statistics pertaining to a group less than some predetermined size $n$ or greater than $N - n$, where $N$ is the total number of records in the database (say). Other techniques, restrict queries with more than some predetermined number of records in common or with too many attributes specified.

Among the techniques that add noise to the statistical results returned are *systematic rounding*, *random rounding*, and *controlled rounding*. The third alternative requires the sum of rounded statistics to equal their rounded sum. The idea is that it is all right if the querier knows the exact answer about a large sample, but nothing should be released about a small sample. Random sample query control is another promising approach to solving the inference problem [**5**]. With this approach each statistic is computed using 80% – 90% of the total number of records and a different sample is used to compute each statistic. For an excellent presentation of these techniques see [**1**].

*Systematic Methods to Enhance Security* – Systematic techniques are used to enhance the security of computer systems. Among these methods are penetration analysis, formal specification and verification, and convert channel analysis. None of these methods guarantees a secure system. They only increase one's confidence in the security of the system.

*Penetration Analysis* – One approach to locating security flaws in computer systems is penetration analysys. This approach uses a collection of known flaws, generalizes these flaws, and tries to apply them to the system being analysed. Usually a team of penetrators, a *team* is given the task of trying to enter the system. Flaws in several major systems have been located using this approach [**6**].

The problem with the *team* approach is that like testing, penetration teams prove the presence, not the absence of protection failures. This observation has led to these of *formal specification* and *verification techniques* to increase ones confidence in the realiability and security of a computer system.

*Formal Verification* – Formal verification demonstrates that an implementation is consistent with the requirements. This task is approached by decomposing it into a number of easier problems. The requirements, which are intially and usually a natural language statement of what is desired, are first stated in precise mathematical terms. This is known as the *formal model* or *criteria* for the system. For example, for a security system the criteria could be that information at one security level doesn't flow into another security level. Next, a high level formal specification of the system is stated. This specification gives a precise mathematical description of the behaviour of the system omitting all implementation details such as recource limitations. This is followed by a series of less abstract specifications each of which implements the next higher level specification, but with more detail. Finally, the system is coded in a higher order language (HOL). This HOL implementation must be shown to be consistent with the original requirements.

It should be emphasized that demonstrating that HOL code is consistent with the security requirements is a difficult process. The process is made tractable by verifying the design at every step. The first step of the verification is to informally verify that the formal model properly reflects the security requirements. This is the only informal step in the process. Since the formal model is at a high level of abstraction and should contain no unnecessary details, it is usually a simple task to review the formal model with the persons who generated the requirements and determine whether the model properly reflects the highest level specifications are consistent with the formal model. Both a state machine approach and an algebraic approach [**8**] are possible.

After the highest level formal specification has been shown to be consistent with the formal model it is necessary to show that the next lower level specification, is one exists, is consistent with the level above it. This process continues from level to level until the lowest level specification is shown to be consistent with the level above it. Finally, it is necessary to show that the HOL implementation is consistent with the lowest level specification. By transitivity, the implementation is thus shown to be consistent with the formal model.

The advent of the security kernel as a means of encapsulating all security relevant aspects of the system makes formal verification feasible. That is, by developing kernel architectures that minimize the amount and complexity of the software involved in security decisions and enforcement, the chances of successfully verifying that the system meets its security requirements are greatly increased. Because the remainder of the system is written using the facilities offered by the kernel, only the kernel code must be verified.

*Covert Channel Analysis* – When performing a security analysis of a system both overt and covert channels of the system must be considered. *Overt channels* use the system's protected data objects to transfer information from one subject to another. That is, one subject writes into a data object and other subject reads from the object; thus, information is transferred from one subject to another. The channels are overt because the entity used to hold the information is a data object; i.e., it is an object that is normally viewed as a data container. Examples of the objects used for overt channels are buffers, files, and devices. *Covert channels* in contrast, use entities not normallly viewed as data objects to transfer information from one device to another. These non-data objects such as file locks, register the state of the system.

Overt channels are controlled by enforcing the access control policy of the system being designed and implemented. Covert channels are more elusive. As stated above, objects used to hold the information being transferred are normally not viewed as data objects, but can often be manipulated maliciously to transfer information. Two types of covert channels are considered in a cover channel analysis: *storage analysis* and *timing channels*. With a storage channel the sending process alters a particular attribute and the receiving attribute to receive information covertly. With a timing channel the sending process modulates the amount of time required for the receiving process to perform a task or detect a change in attribute, and the receiving process interprets this delay or lack of delay as a bit of information.

There are many examples of covert channels and methods for blocking these channels [9]. For example, depleting and restoring resources from a common resource pool is a typical storage channel. A technique commonly used for blocking this type of channel is to have a separate resource pool for each security class and allowing resources to be moved from one pool to another only under the supervision of a flow control method that introduces random delays.

5. **CONCLUSIONS**

This paper has attempted to give a brief overview of the topic of computer security, it has indeed stayed very much to the domain of definitions in this area. A more detailed study of the area is left to the reader. There are many good references on the topic. In particular, [**1 – 10**] provide ample surveys of the area.